%% file: ArticleCDC.tex
\documentclass[10pt,conference]{ieeeconf}

\usepackage{cite}
\usepackage{amsmath,amssymb,amsfonts}
\usepackage{algorithmic}
\usepackage{graphicx}
\usepackage{textcomp}
\usepackage{xcolor}
\usepackage{mathtools}

\usepackage{lipsum}
\usepackage[mathscr]{euscript}
\usepackage{algorithm}
\usepackage{todonotes}

\newtheorem{theorem}{Theorem}
\newtheorem{corollary}[theorem]{Corollary}
\newtheorem{proposition}[theorem]{Proposition}
\newtheorem{definition}[theorem]{\noindent Definition}
\newtheorem{remark}{Remark}
\newtheorem{lemma}[theorem]{Lemma}
\newtheorem{example}{Example}
\newtheorem{assum}{Assumption}

\renewcommand{\QED}{\QEDopen}

\def\mycmd{0} 

\input{commands}

\allowdisplaybreaks

\def\BibTeX{{\rm B\kern-.05em{\sc i\kern-.025em b}\kern-.08em
		T\kern-.1667em\lower.7ex\hbox{E}\kern-.125emX}}

\IEEEoverridecommandlockouts

\begin{document}

\title{Abstracting Linear Stochastic Systems via Knowledge Filtering \thanks{This work is supported by the Dutch NWO Veni project CODEC (project number 18244) and Swedish Research Council International Postdoc Grant 2021-06727. $^1$Department of Electrical Engineering (Control Systems Group), Eindhoven University of Technology, The Netherlands. {$^2$Department of Computer Science, University of Oxford, United Kingdom}  Emails:\{m.h.w.engelaar, m.lazar, s.haesaert\}@tue.nl; \{licio.romao, yulong.gao, alessandro.abate\}@cs.ox.ac.uk}}

\author{M.H.W. Engelaar$^1$, L. Romao$^2$, Y. Gao$^2$ , M. Lazar$^1$,  A. Abate$^2$, and S. Haesaert$^1$}

\maketitle


\begin{abstract}
In this paper, we propose a new model reduction technique for linear stochastic systems that builds upon knowledge filtering and utilizes optimal Kalman filtering techniques. This new technique will reduce the dimension of the noise disturbance and will allow any controller designed for the reduced model to be refined into a controller for the original stochastic system, while preserving any specification on the output. Although initially the reduced model will be time-varying, a method will be provided with which the reduced model can become time-invariant if it satisfies some minor technical conditions. We present our theoretical findings with an example that supports the proposed framework and illustrates how model reduction and controller refinement of stochastic systems can be achieved. We finish the paper by considering specific examples to analyze both completeness with respect to controller synthesis and model order reduction with respect to the state.
%

\end{abstract}


\section{Introduction}

Dynamical systems are becoming more complex, and their tasks more diverse. Adding to this, the inherent uncertainty of most real-life engineering systems \cite{Brec2014,Haes2017_3,Thru2002}, makes developing correct-by-design controllers for stochastic systems, i.e., controllers which ensure the satisfaction of given tasks, an ever-expanding field of research. In the last decade multiple computational tools on correct-by-design control synthesis for stochastic systems have been developed (AMYTISS\cite{Lava2020}, FAU$\text{ST}^2$ \cite{Soud2015}, StocHy \cite{Cauc2019} and SySCoRe \cite{VanH2023}) that can handle complex specifications such as those expressed by temporal languages (LTL, sc-LTL, STL) \cite{Belt2017,Donz2013}. These tools generally suffer from the curse of dimensionality, i.e., exponential growth in computational cost whenever the state-space increases. It has been shown \cite{Haes2017_1,Zama2017} that model reduction can mitigate this effect.
	
The reduction of dynamical models is a mature field in control research. Often known as model order reduction, the objective is to minimize the complexity of a system while retaining some guarantee. For deterministic systems, this includes, amongst others, guarantees on performance \cite{Anto2005}, frequency domain \cite{Guge2008}, and structure \cite{Poly2010} by utilizing, for example, balanced truncation, rational Krylov and moment matching, respectively. Less work exists that can guarantee the satisfaction of complex specifications in the time domain. Examples for deterministic systems include hierarchical control \cite{Gira2009} and simulation relations\cite{Baie2008}. For stochastic systems, this includes such work as (approximate) stochastic simulation relations \cite{Haes2017_1,Haes2017_3}, stochastic simulation functions \cite{Juli2009}, and stochastic bisimulation relations \cite{Pola2017,Zama2014}. These notions allow higher-order (stochastic) systems to be simulated by (finite- or) lower-order systems of (approximately) the same type, all while retaining (approximate) equivalency with regard to their state or output (distribution). For an extensive list of model reduction techniques, see references within \cite{Pola2017}.

Model reduction techniques considering complex temporal specifications are also known as formal abstraction techniques due to frequent usage within formal verification and synthesis. Formal abstraction techniques mainly consists of two components. The first component is the reduction (or abstraction) procedure, which is a procedure that explains how to simplify the original dynamics, thereby obtaining a so-called (simplified) abstract model. The second component is the controller refinement algorithm, which is an algorithm that explains how a correct-by-design controller on the abstract model can be refined into a correct-by-design controller on the original dynamics.

In this paper, we will expand upon the existing literature of formal abstraction by introducing a new model reduction technique that reduces the dimension of the noise disturbance on linear stochastic systems. Similar to the previously mentioned methods, we want to retain the satisfaction of any temporal specification defined on the original system. Accordingly, we will develop a reduction procedure (abstraction procedure) and a controller refinement algorithm. The reduction will consist of two steps: firstly, removing redundant state information, referred to as \emph{knowledge filtering}, which yields a partially observable model, and secondly, computing a reduced realization of this partially observable model via optimal Kalman filtering \cite{Ande2012}. More precisely, the second step implements a weak Gaussian stochastic realization \cite{VanS2021}. We will define a sound controller refinement algorithm and discuss the completeness of the whole approach.

After formalizing the problem statement in Section \ref{Sec:Prel}, a time-varying abstraction will be obtained in Section \ref{Sec:AbsConRefTimVar}.  In Section \ref{Sec:AbsConRefTimInv}, we explain how a time-invariant abstraction can be forced, by considering some technical assumptions. Several examples will be addressed in Section \ref{Sec:CasStu}, illustrating the reduction procedure and controller refinement algorithm. It will also be shown that our method allows for model reduction with respect to the state space where previously mentioned methods are unable. This section will also illustrate a lack of completeness regarding the controller refinement, which can be attributed to the knowledge filtering in the reduction procedure.


\section{Problem Setup} \label{Sec:Prel}

For a given probability measure $\mathbb P$ defined over Borel measurable space $(\X,\mathcal{B}(\X))$, we denote the probability of an event $A\in \Borel(\X)$ as $\mathbb{P}(A)$. In this paper, we will work with Euclidean spaces and Borel measurability.  Details of any measurability considerations are omitted, and we refer the interested reader to \cite{Bert1996}.
\smallskip

\noindent\textbf{Linear Stochastic System. }
We consider a linear time-invariant stochastic system $\M$ given by
\begin{equation} \label{Original}
	\M: \begin{cases}
		x(t+1)	  &= Ax(t) + Bu(t) + w(t)\\
		z(t) 		&= Hx(t),
	\end{cases}
\end{equation}
where $x \in \mathscr{X} \subseteq \mathbb{R}^n$ is the state, $u \in \mathscr{U} \subseteq \mathbb{R}^m$ is the input, $z \in \mathscr{Z} \subseteq \mathbb{R}^p$ is the (performance) output, $x(0) $ is the initial state and is the realization of a Gaussian distribution with mean $\mu_0$ and variance $\Sigma_0$, i.e., $x(0)\sim \mathcal{N}(\mu_0,\Sigma_0)$, and disturbance $w(t)\in \mathbb{R}^n$ is a realization of an independent, identically distributed noise $w(t) \sim \mathcal{N}(0, Q_w)$. Finite executions of $\M$ are alternating sequences of states and inputs ending in a state, such as $\boldsymbol \omega_N^{\text{fin}}=x(0)u(0)x(1)u(1)\dots x(N-1)u(N-1)x(N)$,  which satisfy equation \eqref{Original} for some finite noise sequence $\boldsymbol w = w(0)w(1)w(2)\cdots, w(N-1)$, where $x(0) \sim \mathcal{N}(\mu_0,\Sigma_0)$ and $ w(t)\sim \mathcal{N}(0,Q_w)$ for all $t\in\{0,1,\cdots,N-1\}$. We denote the \textit{history at time $N$} as the set of all finite executions of length $N$ by $\mathcal{E}_N \subseteq (\mathscr{X} \times \mathscr{U})^N \times \mathscr{X}$.

In its most general setting, a controller $\mathbf C$ is a sequence of policies $\C:= \C_0\C_1\C_2\cdots$, such that $\C_t: \mathcal{E}_t \to \mathscr{U}$ is a map of the available history to the set of inputs. The chosen control inputs are given by $u(t)=\C_t(\boldsymbol \omega_t^{\text{fin}})$ and the controlled stochastic system $\C \times \M$ is obtained by composing $\C$ with $\M$. We denote by $\mathcal{Z}:=\mathscr{Z}^{\mathbb{N}}$ the set of all possible output trajectories associated with $\M$. Each execution of $\C \times \M$ will produce an output trajectory $\boldsymbol z = z(0)z(1)z(2) \cdots\in\mathcal{Z}$. The output trajectory $\boldsymbol z$ is a realization of the probability distribution induced by the controlled system $\C \times \M$ and denoted as $\boldsymbol z \sim \mathbb{P}_{\C \times \M}$. 
\smallskip


\noindent \textbf{Stochastic Correct-by-Design Control Synthesis. }
Let us consider the goal of designing a controller $\C$ that ensures output trajectories of the controlled system $\C \times \M$ satisfy a given specification $\phi$. We assume that each specification $\phi$ corresponds to a Borel measurable subset of $\mathcal{Z}$, denoted by $\mathcal{Z}_\phi \in \B(\mathcal{Z})$. Examples of such specifications  include specifications given in linear-time temporal logics \cite{Baie2008}. Given the stochastic nature of $\M$, it is natural to require that the specification $\phi$ is satisfied by the controlled system $\C \times \M$, with probability at least $p$. Let us denote the satisfaction probability as $\mathbb{P}_{\C \times \M}(\mathcal{Z}_\phi)$. Then, the objective is to synthesize $\C$ such that $\mathbb{P}_{\C \times \M}(\mathcal{Z}_\phi)\geq p$.  We refer to this as \emph{stochastic correct-by-design control synthesis}.

\smallskip


\noindent\textbf{Problem statement. }
To mitigate scaling issues such as the curse of dimensionality in stochastic correct-by-design control synthesis, we are interested in designing an abstract model for which the stochastic control synthesis problem is substantially simpler while also preserving correctness to specifications defined on the output trajectories. More precisely, our goal is to construct a \emph{noise reduced} abstract model $\bar \M$ such that for any correct-by-design controller $\bar \C$ synthesized for the abstract model $\bar \M$, a correct-by-design controller $\C$ can be obtained for the original model $\M$ with equal satisfaction probability, i.e.,
\begin{equation} \label{StochProb}
	\forall \bar \C  \ \exists \C: \Pr_{\bar \C \times \bar \M}(\mathcal{Z}_\phi)  = \Pr_{\C \times \M}(\mathcal{Z}_\phi).
\end{equation}
In the remainder, we constructively solve this problem.

\section{Abstraction and Control Refinement: Time-varying Abstraction} \label{Sec:AbsConRefTimVar}

\begin{figure}[htp]
	\includegraphics[trim={6cm 0 6cm 0},clip,width=0.96\columnwidth]{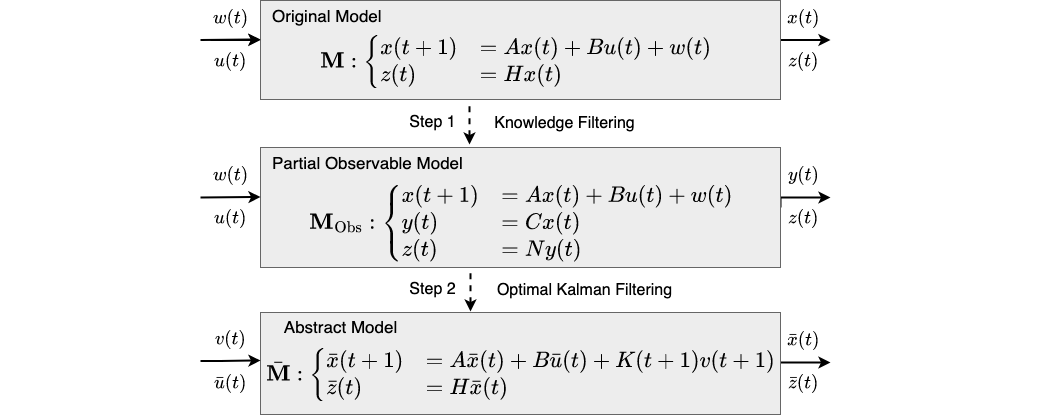}
	\centering
	\caption{The block diagram illustrates the abstraction procedure. The distribution of $v(t)$ depends on the distribution of $y(t)$. $K(t)$ is the time-varying Kalman gain obtained from the Kalman filter equations.\vspace{-1.2em}}
\label{Fig:Proc}
\end{figure}

Consider a stochastic system $\M$ as given by \eqref{Original}. To solve the stochastic correct-by-design control synthesis problem in \eqref{StochProb} while also simplifying the noise, we introduce the abstraction procedure illustrated in Fig. \ref{Fig:Proc} and the controller refinement algorithm represented in Algorithm \ref{Alg:ConRef}.

The abstraction procedure hinges on removing potential redundant information from the original model $\M$ without influencing the performance output $\boldsymbol z$. The procedure is executed in two steps. The first step filters knowledge from $\M$ by introducing an observation output $y(t) = Cx(t)$. The result is a new stochastic system $\M_{\text{Obs}}$ that is partially observable, or more specific, a partially observable Markov decision process \cite{Kris2016}. The second step replaces the partially observable model with a fully observable equivalent model via optimal Kalman filtering \cite{Ande2012}. We refer to this fully observable model as the abstract model $\bar \M$. 


In the following, we will elucidate these steps for a time-varying abstract model and show that the controller refinement algorithm is valid. In the next section, we present some technical conditions under which a time-invariant abstract model can be obtained.\\[-1.7em]


\subsection{Abstract Model Construction} \label{Sec:AbsModCon}

Following the steps of the abstraction procedure Fig. \ref{Fig:Proc}, we will construct the abstract model $\bar\M$. For the first step, we  choose a matrix pair $C$ and $N$ with $C \in \mathbb{R}^{q \times n}$ and $ N \in \mathbb{R}^{p \times q}$, such that $NC=H$ and $q < n$, and define the partial observable version of the original model as 
\begin{equation} \label{OriginalPOMDP}
	\M_{\text{Obs}}: \begin{cases}
		 x(t+1)	  &= A x(t) + B u(t) +  w(t)\\
		 y(t)		&= C  x(t)\\
		 z(t)		&= N  y(t),
	\end{cases}
\end{equation}
where $ x(0) \sim \mathcal{N}(\mu_0,\Sigma_0)$, $y(t) \in \mathbb{R}^q$ is the observation output, and $ w(t) \sim \mathcal{N}(0,Q_w)$. For the second step, we first introduce the needed optimal Kalman filtering techniques for estimating the state of \eqref{OriginalPOMDP} based on observations and inputs. We denote by $x_K(t {\mid} t)$ the expectation of $x(t)$, conditional on the available information at time $t$, that is, $x_K(t {\mid} t)= \mathbb{E} (x(t)|y(0)u(0)...u(t-1)y(t))$. Similarly, we denote by $P(t {\mid} t)$ the variance of $x(t)$, again conditional on the available information at time $t$, i.e., $P(t {\mid} t)=\text{Var}(x(t) {\mid} y(0)u(0)\cdots y(t))$. Quantities $x_K(t {\mid} t)$ and $P(t {\mid} t)$ are called, respectively, \emph{a posteriori state estimate} and \emph{a posteriori state variance}. The term a posteriori is used because all available information, including $y(t)$, is utilized in both definitions of $x_K(t {\mid} t)$ and $P(t {\mid} t)$. The \emph{a priori} quantities $x_K(t {\mid} t-1)$ and $P(t {\mid} t-1)$ are defined similarly, but contrary to the a posteriori quantities, only consider the available information up till $u(t-1)$. As is common, these quantities can be determined as follows
\begin{subequations} \label{KalFilEq}
	\begin{align}
		x_K(t {\mid} t) &= x_K(t {\mid} t-1) \nonumber \\
		&\quad + K(t)[ y(t)-C x_K(t {\mid} t-1)], \label{KalMeaUpdSta}\\
		x_K(t+1 {\mid} t) &= Ax_K(t {\mid} t) + B u(t), \label{KalTimUpdSta}\\
		P(t {\mid} t)&= P(t {\mid} t-1) - K(t)CP(t {\mid} t-1), \label{KalMeaUpdCov} \\
		P(t+1 {\mid} t) &= A P(t {\mid} t) A^T + Q_w, \label{KalTimUpdCov} \\
		K(t) &= \ P(t {\mid} t-1)C^T[CP(t {\mid} t-1)C^T]^{-1}, \label{KalFilGai}\\
		x_K(0 {\mid}-1) &=\mu_0 \text{ and } P(0 {\mid} -1) =  \Sigma_0. \label{KalFilIni}
	\end{align}
\end{subequations}
We will make the following assumption to ensure equations \eqref{KalFilEq} are valid throughout this section.
\begin{assum} \label{Assump}
	$C\Sigma_0C^T  \succ 0$ and $CQ_wC^T  \succ 0$, i.e., both are strictly positive definite.
\end{assum}

It is well known that the error between the a priori prediction of the output, $Cx_K(t {\mid} t-1)$, and the measured output $y(t)$ defines a Gaussian white noise sequence \cite{VanS2021}. This noise sequence is referred to as the innovation and defined by
\begin{equation} \label{Eq:Inn}
	v(t) =  y(t) - C x_K(t {\mid} t-1).
\end{equation}
The mean and variance of $v$ can be computed directly, giving $\mu_{v}(t) = 0$ and $\Sigma_{v}(t) = CP(t {\mid} t-1)C^T$, see also \cite[Theorem 14.4.2]{VanS2021}. The innovation allows us to define a new linear stochastic system. This new system, referred to as the \emph{a priori} innovation process, builds on the a priori quantities $\hat x(t)=x_K(t|t-1)$ and $\hat P(t) =P(t|t-1)$ and is given by
\begin{equation} \label{InnPro}
	\hat \M :\begin{cases}
		\hat x(t+1) &= A\hat x(t) + Bu(t) + AK(t)v(t)\\
		y(t) &= C \hat x(t) +v(t)\\
		\hat z(t) &= Ny(t),
	\end{cases}
\end{equation}
\\[-2.4em]
\begin{align*}
 \mbox{where }	\hat x(0) &= \mu_0, \  v(t) \sim \mathcal{N}(0,\Sigma_{v}(t)), \ \hat P(0) = \Sigma_0\\
	\hat P(t+1) &= A \hat P(t)A^T+ Q_w-AK(t) C\hat P(t)A^T,\\
	K(t) &= \hat P(t)C^T[C\hat P(t)C^T]^{-1}, \ \Sigma_{v}(t) = C\hat P(t)C^T.
\end{align*}

Although this is the most commonly used version of the innovation process, we will now define an alternative one based on the a posteriori quantities $\bar x(t)=x_K(t|t)$ and $\bar P(t) =P(t|t)$. We refer to this as the \emph{a posteriori} innovation process, and this will be our abstract model $\bar \M$, defined as
\begin{equation}\label{CurInnPro}
	\bar \M:
	\begin{cases}
		\bar x(t+1)	  &\!\!\!\!= A\bar x(t) + Bu(t) +K(t+1) v(t+1) \\
		y(t) & \!\!\!\!= C\bar x(t)\\
		\bar z(t) & \!\!\!\! = Ny(t),
	\end{cases}
\end{equation}
\\[-2.1em]
\begin{subequations} \label{AbsCon}
\begin{align}
\mbox{\hspace{-2.0em}where\quad }	\bar x(0) &\sim \ \mathcal{N}(\mu_0, \Sigma_0-\bar P(0)),\\
	v(t) &\sim \mathcal{N}(0,\Sigma_{v}(t)),\ \Sigma_{v}(t) = C\hat P(t)C^T,\label{eq:noise_dist}\\
	\bar P(t) &= \hat{P}(t) -K(t) C\hat P(t),\\
	\hat P(t+1) &= A \bar{P}(t) A^T + Q_w,\\
	K(t) &= \hat P(t)C^T[C\hat P(t)C^T]^{-1},\\
	\bar P(0) &= \Sigma_0 - \Sigma_0 C^T[C \Sigma_0 C^T]^{-1}C\Sigma_0.
\end{align}
\end{subequations}
\if\mycmd0
In the appendix,
\else
In the appendix of the extended version of this paper \cite{Enge2023},
\fi
additional information is given concerning the derivation of both innovation processes. Finally, the definitive definition of the abstract model is given by
\begin{equation}\label{Abstract}
	\bar \M:
	\begin{cases}
		\bar x(t+1)	  &\!\!\!\!= A\bar x(t) + B\bar u(t) +K(t+1) v(t+1) \\
		\bar z(t) & \!\!\!\!= H\bar x(t),
	\end{cases}
\end{equation}
together with the equations \eqref{AbsCon}. The abstract model $\bar \M$ is a \emph{weak Gaussian stochastic realization} of $\M_{\text{Obs}}$  \eqref{OriginalPOMDP} as defined in\cite{VanS2021}. By construction, M and $\bar \M$ have the same state-space dimension; however, the noise affecting the latter takes value in an Euclidean space of smaller dimension, due to the knowledge filtering in step 1.
More precisely, the noise input $w(t)\sim\mathcal N(0,Q_w)$ with $w(t)\in\mathbb R^n$ has been replaced with the noise input $K(t+1)v(t+1)$ with $v(t+1)\in \mathbb R^q$, where $q < n$. This reduction in complexity comes at the cost of having a time-varying system. In the next section, we present conditions under which the resulting $\bar \M$ is time-invariant.


\subsection{Controller Refinement} \label{Sec:ConRefVar}

\begin{figure}[H]
	\includegraphics[trim={1.7cm 0 -1cm 0},clip,width=0.95\columnwidth]{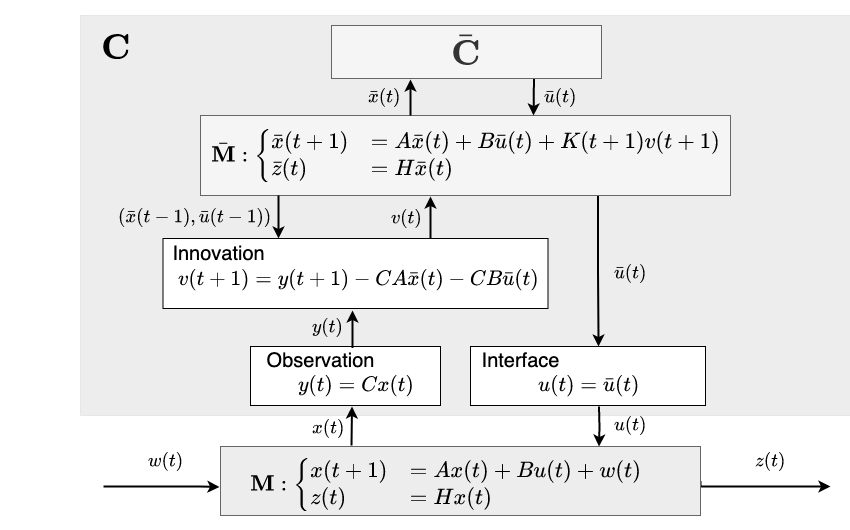}
	\centering
	\vspace{-0.5em}
	\caption{A block diagram for the controller refinement (Algorithm \ref{Alg:ConRef})}\vspace{-1em}
	\label{Fig:ConRef}
\end{figure}

What remains to be shown is that for any correct-by-design controller $\bar\C$ designed for the abstract model, there exists a correct-by-design controller $\C$ for the original model, see equation \eqref{StochProb}. We use the auxiliary output $y(t)=Cx(t)$ to define the following controller refinement Algorithm \ref{Alg:ConRef} -- its implementation is depicted in the block diagram of Fig. \ref{Fig:ConRef}.
\begin{algorithm}[H]
	\caption{Controller refinement algorithm}
	\begin{algorithmic}[1] \label{Alg:ConRef}
		\STATE Given: $\M$, $\bar \M$, $\bar \C$
		\STATE set $t:=0$ and compute
		$K(0):=\Sigma_0C^T(C\Sigma_0C^T)^{-1}$,
		\STATE draw $x(0)$ from $\mathcal{N}(\mu_0,\Sigma_0)$,
		\STATE compute $\bar x(0)=\mu_0+K(0)(Cx(0)-C\mu_0)$,
		\LOOP
		\STATE obtain $\bar u(t)$ according to $\bar \C$,
		\STATE set $u(t)=\bar u(t)$, \COMMENT{ $\leftarrow$ Implementing $\C$}
		\STATE draw $x(t+1)$ from $\M$ and get $y(t+1)=Cx(t+1)$,
		\STATE  compute $v(t+1)=y(t+1)-CA\bar x(t)-CB \bar u(t)$,
		\STATE 	compute $\bar x(t+1) = A \bar x(t) + B \bar u(t) +K(t+1) v(t+1)$
		\STATE take $t=t+1$.
		\ENDLOOP
	\end{algorithmic}
\end{algorithm}
The following theorem constitutes one of the main contributions of this paper.

\begin{theorem}\label{Thm:main}
	Consider $\M$ and $\bar \M$ as given by, respectively, \eqref{Original} and \eqref{Abstract} and assume that $C\Sigma_0C^T  \succ 0$ and $CQ_wC^T  \succ 0$. Let $\phi$ be any specification described by $\mathcal{Z}_{\phi} \in \B(\mathcal{Z})$. Then $\forall \bar \C \ \exists \C$ such that $\Pr_{\bar \C \times \bar \M}(\mathcal{Z}_\phi)=\Pr_{\C \times \M}(\mathcal{Z}_\phi)$.
\end{theorem}
\if\mycmd0
\begin{proof}
	Consider a controller $\C$ constructed based on Algorithm \ref{Alg:ConRef}. Using standard optimal Kalman filtering arguments, one can show that, at every time step, the realization of $v(t+1)$ in Algorithm \ref{Alg:ConRef} is uncorrelated with the current state $\bar x(t+1)$ and has distribution given by \eqref{eq:noise_dist}. Therefore, the embedded model $\bar \M$ has the same output distribution as $\bar \M$ in \eqref{Abstract} when both are given the same input sequence. Similarly, we claim that model $\bar \M$ in \eqref{Abstract} and $\M$ have the same output distribution when given the same input sequence. The proof of this claim can be found in the appendix. Hence, the embedded model $\bar \M$ has the same output distribution as $\M$ when both are given the same input sequence. This finishes the proof, as $\Pr_{\bar \C \times \bar \M}(\mathcal{Z}_\phi)= p$, now implies $\Pr_{\C \times \M}(\mathcal{Z}_\phi)= p$.
\end{proof}
\else
\begin{proof}
	Consider a controller $\C$ constructed based on Algorithm \ref{Alg:ConRef}. Using standard optimal Kalman filtering arguments, one can show that, at every time step, the realization of $v(t+1)$ in Algorithm \ref{Alg:ConRef} is uncorrelated with the current state $\bar x(t+1)$ and has distribution given by \eqref{eq:noise_dist}. Therefore, the embedded model $\bar \M$ has the same output distribution as $\bar \M$ in \eqref{Abstract} when both are given the same input sequence. Similarly, we claim that model $\bar \M$ in \eqref{Abstract} and $\M$ have the same output distribution when given the same input sequence. The proof of this claim can be found in the appendix of the extended paper \cite{Enge2023}. Hence, the embedded model $\bar \M$ has the same output distribution as $\M$ when both are given the same input sequence. This finishes the proof, as $\Pr_{\bar \C \times \bar \M}(\mathcal{Z}_\phi)= p$, now implies $\Pr_{\C \times \M}(\mathcal{Z}_\phi)= p$.
\end{proof}\fi


\section{Abstraction and Control Refinement: Time-invariant Case}\label{Sec:AbsConRefTimInv}

In this section, we investigate sufficient conditions to obtain a time-invariant abstract model based on the abstraction procedure explained in Section \ref{Sec:AbsModCon}. From the block diagram in Fig. \ref{Fig:ConRef}, one may deduce that a time-invariant model is derived if, amongst others, a time-invariant Kalman gain is employed.



\subsection{Abstract Model Construction} \label{Sec:AbsModConInv}

As explained in the sequel, towards obtaining a time-invariant abstract model $\bar \M$, we need to define the following algebraic equation.
\begin{definition}[Discrete Algebraic Ricatti Equation \cite{Ione1993}]
	The discrete-time algebraic Ricatti equation (DARE) adapted to the model \eqref{OriginalPOMDP} is given by\\[-0.9em]
	\begin{equation} \label{Eq:DARE}
		X=AXA^T-AXC^T[CXC^T]^{-1}CXA^T+Q_w.
	\end{equation}
	We say that $X \succ 0$ is a stabilizing solution to the DARE, if $A-FC$ is stable for $F=AXC^T(CXC^T)^{-1}$. 
\end{definition}
In case the variance of the initial condition (denoted by $\Sigma_0$) is a stabilizing solution to the algebraic Ricatti equation in \eqref{Eq:DARE}, we have that $\hat{P}(t) = \Sigma_0$ in \eqref{AbsCon}, i.e., the a priori state variance becomes constant. As a result the abstract model \eqref{Abstract} will be time-invariant, which is formalized by the following lemma.
\begin{lemma} \label{Lem:WeakTimInv}
	Assume that $C\Sigma_0C^T \succ 0$. If $\Sigma_0=X \succ 0$ is a stabilizing solution to \eqref{Eq:DARE}, then the abstract model \eqref{Abstract} is time-invariant and given by
	\begin{equation}\label{AbstractInv}
		\bar \M:
		\begin{cases}
			\bar x(t+1)	  &= A\bar x(t) + B\bar u(t) +Kv(t+1) \\
			\bar z(t) & = H\bar x(t)
		\end{cases}
	\end{equation}
	\\[-2.6em]
	\begin{align*}
		\mbox{where }\bar x(0) &\sim \ \mathcal{N}(\mu_0,X-P), \ 	\bar v(t)  \sim \mathcal{N}(0,CXC^T),\\
		K &= XC^T[CXC^T]^{-1}, \ P = X - KCX.
	\end{align*}
\end{lemma}

The main advantage of Lemma \ref{Lem:WeakTimInv} is the guarantee of the abstract model \eqref{AbstractInv} being time-invariant, contrary to the abstract model \eqref{Abstract}, which may be time-varying. Regrettably, most real-life engineering systems do not yield the result of Lemma \ref{Lem:WeakTimInv}, as, generally, $\Sigma_0$ will not solve \eqref{Eq:DARE}. To alleviate this issue, we will consider a relaxed version that requires instead that $\Sigma_0-X \succ 0$, with $X \succ 0$ being the stabilizing solution to \eqref{Eq:DARE}. Accordingly, we have the following assumption for the remainder of this subsection.
\begin{assum}
	$CXC^T \succ 0$ and $\Sigma_0-X \succ 0$ with $X \succ 0$ being the stabilizing solution to \eqref{Eq:DARE}.
\end{assum}

To obtain a time-invariant abstract model based on the above assumption, we again consider the abstraction procedure explained in Section \ref{Sec:AbsModCon}. While step 1 remains the same, step 2 will be changed slightly. An additional observation $\tilde y=x(0)+\tilde w$, where $\tilde w \sim \mathcal{N}(0,R)$, will be added at time $t=0$. This will allow for modification of the a posteriori quantities of $x(0)$, $x_K(0|0)$ and $P(0|0)$, before continuing the abstraction procedure by utilizing the Kalman filter equations \eqref{KalFilEq}, excluding \eqref{KalFilIni}. The goal is to ensure that $P(0|0)=X-XC^T[CXC^T]^{-1}CX$ implying that $P(t+1|t)=X$, $\forall t \in \{0,1,2,\dots \}$ in \eqref{KalFilEq}. This will make the a posteriori innovation process time-invariant. To accomplish this, we take $R = (X^{-1}-\Sigma_0^{-1})^{-1}$. The result will be an abstract model $\bar \M^*$, given by
\begin{equation}\label{AbstractInvStr}
	\bar \M^*:
	\begin{cases}
		\bar x(t+1)	  &= A\bar x(t) + B\bar u(t) +K v(t+1) \\
		\bar z(t) & = H\bar x(t)
	\end{cases}
\end{equation}
\begin{align*}
\mbox{where } \bar x(0) &\sim \mathcal{N}(\mu_0,\Sigma_0-P), \ v(t) \sim \mathcal{N}(0,CXC^T)\\
	K &= XC^T[CXC^T]^{-1}, \,
	P = X-KCX.
\end{align*}
\if\mycmd0
Note that $\bar \M^*$ differs from the abstract model \eqref{AbstractInv} only in the initial distribution. See the appendix for a more detailed explanation on how to obtain $\bar \M^*$.
\else
Note that $\bar \M^*$ differs from the abstract model \eqref{AbstractInv} only in the initial distribution. See the appendix of the extended paper \cite{Enge2023} for a more detailed explanation on how to obtain $\bar \M^*$.\fi
	


\subsection{Controller Refinement}

Under the conditions of Lemma \ref{Lem:WeakTimInv}, Algorithm \ref{Alg:ConRef} will again give a valid controller refinement algorithm and Theorem \ref{Thm:main} can be rephrased as follows.
\begin{corollary}\label{Cor:main}
	Consider $\M$ and $\bar \M$ as given by, respectively, \eqref{Original} and \eqref{AbstractInv} and assume that $\Sigma_0 \succ 0$ is a stabilizing solution to \eqref{Eq:DARE}, and $C\Sigma_0C^T \succ 0$. Let $\phi$ be any specification described by $\mathcal{Z}_{\phi} \in \B(\mathcal{Z})$. Then $\forall \bar \C \ \exists \C$ such that $\Pr_{\bar \C \times \bar \M}(\mathcal{Z}_\phi)=\Pr_{\C \times \M}(\mathcal{Z}_\phi)$.
\end{corollary}

Should instead abstract model $\bar \M^*$ be considered, Algorithm \ref{Alg:ConRef} needs to be slightly modified, resulting in Algorithm \ref{Alg:ConRefInv}. Note that Algorithm \ref{Alg:ConRefInv} uses an auxiliary step to ensure the initialization is resolved correctly.\\[-1.5em]
\begin{algorithm}[htp]
	\caption{Controller refinement algorithm.}
	\begin{algorithmic}[1] \label{Alg:ConRefInv}
		\STATE Given: $\M$, $\bar \M^*$, $\bar \C^*$
		\STATE set $t:=0$ and compute $L=\Sigma_0[\Sigma_0+R]^{-1}$,
		\STATE draw $x(0)$ from $\mathcal{N}(\mu_0,\Sigma_0)$ and draw $\tilde w$ from $\mathcal{N}(0,R)$,
		\STATE compute $\bar \mu_0= \mu_0+L(x(0)+\tilde w-\mu_0)$,
		\STATE compute $\bar x(0)=\bar \mu_0+K(Cx(0)-C\bar \mu_0)$,
		\LOOP
		\STATE obtain $\bar u(t)$ according to $\bar \C^*$,
		\STATE set $u(t)=\bar u(t)$, \COMMENT{ $\leftarrow$ Implementing $\C$}
		\STATE draw $x(t+1)$ from $\M$ and get $y(t+1)=Cx(t+1)$,
		\STATE 
		compute $v(t+1)=y(t+1)-CA\bar x(t)-CB \bar u(t)$,
		\STATE 
		compute $\bar x(t+1) = A \bar x(t) + B \bar u(t) +K v(t+1)$
		\STATE take $t=t+1$.
		\ENDLOOP
	\end{algorithmic}
\end{algorithm}
\\[-2em]

Based on Algorithm \ref{Alg:ConRefInv}, we can now extend the result of Theorem \ref{Thm:main} to the abstract model $\bar \M^*$.
\begin{theorem} \label{Thm:Inv}
	Consider $\M$ and $\bar \M^*$ as given by, respectively, \eqref{Original} and \eqref{AbstractInvStr}. Let $X \succ 0$ be a stabilizing solution to \eqref{Eq:DARE}, and assume that $\Sigma_0-X \succ 0$ and $CXC^T \succ 0$. Let $\phi$ be any specification described by $\mathcal{Z}_{\phi} \in \B(\mathcal{Z})$. Then $\forall \bar \C^* \ \exists \C$ such that $\Pr_{\bar \C^* \times \bar \M^*}(\mathcal{Z}_\phi)=\Pr_{\C \times \M}(\mathcal{Z}_\phi)$.
\end{theorem}
\begin{proof}
	The proof follows from Algorithm \ref{Alg:ConRefInv}, similar to Theorem \ref{Thm:main} only now with two initial measurements.
\end{proof}

\begin{remark}
	Due to the knowledge filtering in step 1 of the abstraction procedure, in general, Theorem \ref{Thm:main}, Corollary \ref{Cor:main} and Theorem \ref{Thm:Inv} do not hold when reversing the statement, that is, the existence of $\C$ such that $\Pr_{\C \times \M}(\mathcal{Z}_\phi)= p$ does not imply existence of $\bar \C$ such that $\Pr_{\bar \C \times \bar \M}(\mathcal{Z}_\phi)= p$. This will be further illustrated in the following section.
\end{remark}


\section{Stochastic Correct-by-Design Control Synthesis: Examples} \label{Sec:CasStu}

In this section, we will consider an example to illustrate the abstraction procedure and the controller refinement algorithm. Another example will show that model reduction with respect to the state can be achieved, under the right conditions, where previous existing methods are inadequate. Finally, we will illustrate that, by filtering knowledge, we may construct an abstract model for which synthesis of a correct-by-design controller is not possible.

\begin{figure}[htp]
	\includegraphics[trim={3cm 0cm 2cm 0cm},clip,width=\columnwidth]{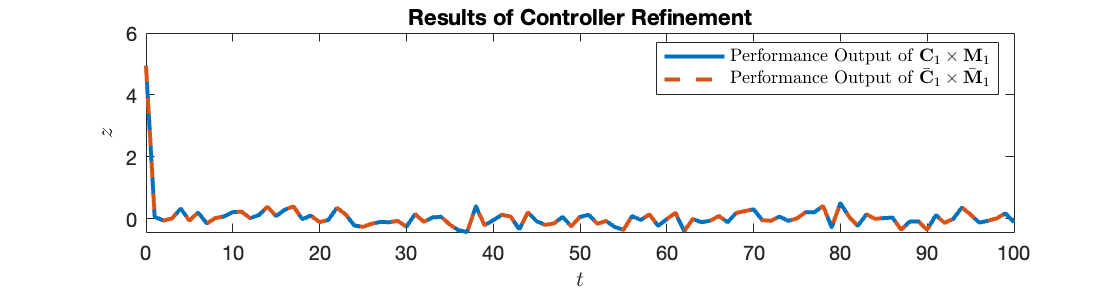}
	\centering
	\vspace{-2em}
	\caption{The performance output of the original- and embedded abstract model over a horizon $t \in [0,100]$, when applying Algorithm \ref{Alg:ConRefInv}.}
	\label{Fig:Example1}
\end{figure}

\begin{example} \label{Exa:1}
	Consider the discrete-time stochastic system
	\begin{align}\label{eq:Mexample}
		\M: &\begin{cases}
			x(t+1) &\!\!\!\!\!=\!\! \begin{bmatrix}
				0 & 0 & 0\\
				1 & 0 & 0\\
				0 & 1 & 0
			\end{bmatrix} x(t) +
			\begin{bmatrix}
				0\\
				0\\
				1
			\end{bmatrix} u(t) + w(t)\\
			z(t) &\!\!\!\!\!=\!\! \begin{bmatrix}
				0 & 0 & 1
			\end{bmatrix} x(t),
		\end{cases}
	\end{align}
	where $x(0) \sim \mathcal{N}(0,\Sigma_0)$ and $w \sim \mathcal{N}(0,Q_w)$ with
	\begin{align}\notag
		\Sigma_0 = \begin{bsmallmatrix}
			5 & 0 & 0\\
			0 & 5 & 0\\
			0 & 0 & 5\\
		\end{bsmallmatrix} \text{ and } Q_w &= \begin{bsmallmatrix}
			1 & 0 & 0\\
			0 & 1 & 0\\
			0 & 0 & 0.05\\
		\end{bsmallmatrix}. 
	\end{align}
	Let $\phi$ be a temporal specification, which requires $z$ to be within $[-1,1]$ over the interval $[1,100]$. We aim to design a controller $\C$ such that $\Pr_{\C \times \M}(\mathcal{Z}_\phi) \geq 0.95$.
	
	Let us construct an abstract model $\bar \M_{1}$ utilizing information from the second and third state, that is, let $C_1=\begin{bmatrix} 0 & 1 & 0\\ 0 & 0 & 1 \end{bmatrix}$ and $N_1=\begin{bmatrix} 0 & 1 \end{bmatrix}$, and notice that $[0~0~1] = N_1 C_1$, satisfying the condition $H=NC$. Let $X = \begin{bsmallmatrix}
		1 & 0 & 0 \\ 0 & 2 & 0 \\ 0 & 0 & 0.05
	\end{bsmallmatrix}$ be the solution to \eqref{Eq:DARE} associated with $\M$, and observe that  $\Sigma_0- X \succ 0$ and $ C_1XC_1^T \succ 0$. The abstract model is obtained from \eqref{AbstractInvStr} and given by
	\begin{align*}
		\bar \M_{1}	: &\begin{cases}
			\bar x_1(t+1) &\!\!= \begin{bsmallmatrix}
				0 & 0 & 0\\
				1 & 0 & 0\\
				0 & 1 & 0
			\end{bsmallmatrix} \bar x_1(t) +
			\begin{bsmallmatrix}
				0\\
				0\\
				1
			\end{bsmallmatrix} \bar u_1(t) +
			\begin{bsmallmatrix}
				0 & 0 \\
				1 & 0 \\
				0 & 1
			\end{bsmallmatrix}
			v_1(t)\\
			\bar z_1(t) &\!\!= \begin{bsmallmatrix}
				0 & 0 & 1
			\end{bsmallmatrix} \bar x_1(t)
		\end{cases}
	\end{align*}
	where $\bar x_1(0) \sim \mathcal{N}(0,\begin{bsmallmatrix} 4 & 0 & 0\\ 0 & 5 & 0 \\ 0 & 0 & 5 \end{bsmallmatrix})$ and $v_1 \sim \mathcal{N}(0,\begin{bsmallmatrix}
		2 & 0\\
		0 & 0.05\\
	\end{bsmallmatrix})$. Notice that this abstract model is time-invariant. 
	
	For $\bar \M_1$ to satisfy the specification, take $\bar \C_1: \bar u_1(t)=\begin{bmatrix} 0 & -1 & 0 \end{bmatrix} \bar x_1(t).$ The result will be that $\bar z_1(t)=\begin{bmatrix} 0 & 1 \end{bmatrix}v_1(t+1)$, that is, $\bar z_1(t) \sim \mathcal{N}(0,0.05)$. Using the cumulative distribution function of $\mathcal{N}(0,0.05)$, we can compute that $\Pr(\bar z_1(t) \notin [-1,1]) = 7.744\text{e-}6$ for all $t\in [1,100]$. This implies that  $\Pr(\bar{\boldsymbol{z}}_1 \notin \mathcal{Z}_\phi)=7.741\text{e-}4$ and $\Pr_{\bar \C_1 \times \bar \M_1}(\mathcal{Z}_\phi) > 0.95$. To obtain controller $\C_1$, we utilize Algorithm \ref{Alg:ConRefInv}. In Fig. \ref{Fig:Example1}, the result of applying Algorithm \ref{Alg:ConRefInv} is shown. \QED\vspace{-1em}
\end{example}
\vspace{-0.9em}

The above example illustrates that the proposed abstraction procedure yields a simplified system in regard to the noise and the controller refinement algorithm will produce a correct-by-design controller for the original model. Even though we considered a simple specification in this example, these results remain unaltered for more complex properties. Similarly, a trivial control design was used in this example, but should any other correct-by-design controller on the abstract model be used, these results remain again unaltered.
\vspace{-0.9em}



\noindent{\textbf{Model Reduction with respect to the State.}}
We now investigate model reduction with respect to the state.

\begin{example}[continued from Ex. \ref{Exa:1}] \label{Exa:2}
	Assume $\Sigma_0=X$, a stabilizing solution to \eqref{Eq:DARE}. Inspired by Lemma \ref{Lem:WeakTimInv}, consider the abstract model
	\begin{align*}
		\bar \M_{2}	: &\begin{cases}
			\bar x_2(t+1) &\!\!= \begin{bsmallmatrix}
				0 & 0 & 0\\
				1 & 0 & 0\\
				0 & 1 & 0
			\end{bsmallmatrix} \bar x_2(t) +
			\begin{bsmallmatrix}
				0\\
				0\\
				1
			\end{bsmallmatrix} \bar u_2(t) +
			\begin{bsmallmatrix}
				0 & 0 \\
				1 & 0 \\
				0 & 1
			\end{bsmallmatrix}
			v_2(t)\\
			\bar z_2(t) &\!\!= \begin{bsmallmatrix}
				0 & 0 & 1
			\end{bsmallmatrix} \bar x_2(t)
		\end{cases}
	\end{align*}
	where $\bar x_2(0) \sim \mathcal{N}(0,\begin{bsmallmatrix} 0 & 0 & 0\\ 0 & 2 & 0 \\ 0 & 0 & 0.05 \end{bsmallmatrix})$ and $v_2 \sim \mathcal{N}(0,\begin{bsmallmatrix}
		2 & 0\\
		0 & 0.05\\
	\end{bsmallmatrix})$. 
	Due to its structure, system $\bar \M_{2}$ can be reduced to 
	\begin{align*}\label{eq:Mexred2}
		\bar \M_{2,r}	: &\begin{cases}
			\bar x_2(t+1) &= \begin{bsmallmatrix}
				0 & 0\\
				1 & 0
			\end{bsmallmatrix} \bar x_2(t) +
			\begin{bsmallmatrix}
				0\\
				1
			\end{bsmallmatrix} \bar u_2(t) +
			v_2(t)\\
			\bar z_2(t) &= \begin{bsmallmatrix}
				0 & 1
			\end{bsmallmatrix} \bar x_2(t)
		\end{cases}
	\end{align*}
	where $\bar x_2(0) \sim \mathcal{N}(0,\begin{bsmallmatrix} 2 & 0 \\ 0 & 0.05 \end{bsmallmatrix})$ and $v_2 \sim \mathcal{N}(0,\begin{bsmallmatrix}
		2 & 0\\
		0 & 0.05\\
	\end{bsmallmatrix})$. 
	Utilizing the lower dimensional system  $\bar \M_{2,r}$, we obtain the correct-by-design controller $\bar \C_2: \bar u_2(t)=\begin{bmatrix} -1 & 0 \end{bmatrix} \bar x_2(t)$, after which we use Algorithm \ref{Alg:ConRef} to obtain controller $\C_2$. \QED
\end{example}

It is important to note that the above reduction cannot be quantified by existing simulation relations such as \cite{Haes2017_1,Juli2009,Pola2017}. This makes our method a promising new model reduction technique, but for which further research is still necessary. Notice that in Algorithm \ref{Alg:ConRef}, a minor modification needs to be made based on the obtained reduced model. For instance, in the above example, one must remove the first element of $\bar x(t)$ before feeding the remainder to $\bar \C_2$.
\smallskip


\noindent{\textbf{Lack of Completeness.}}
The proposed framework is not complete. Due to our choice of knowledge filtering, the original problem might become oversimplified to such a degree that no correct-by-design controller can be obtained for the abstract model. However, this failure to design a controller using the abstract model does not imply that the specification cannot be enforced on the original dynamics.

\begin{example}[continued from Ex. \ref{Exa:1}] \label{Exa:3}
	Consider again the linear time-invariant stochastic system \eqref{eq:Mexample} with constraint $\Pr_{\C \times \M}(\mathcal{Z}_\phi) \geq 0.95$. Consider now matrices $C_2=\begin{bmatrix} 0 & 0 & 1 \end{bmatrix} \text{ and } N_2= 1$,
	which leads to the following abstract model
	\begin{align*}
		\bar \M_{3}	: &\begin{cases}
			\bar x_3(t+1) &= \begin{bsmallmatrix}
				0 & 0 & 0\\
				1 & 0 & 0\\
				0 & 1 & 0
			\end{bsmallmatrix} \bar x_3(t) +
			\begin{bsmallmatrix}
				0\\
				0\\
				1
			\end{bsmallmatrix} \bar u_3(t) +
			\begin{bsmallmatrix}
				0 \\
				0 \\
				1
			\end{bsmallmatrix}
			v_3(t)\\
			\bar z_3(t) &= \begin{bsmallmatrix}
				0 & 0 & 1
			\end{bsmallmatrix} \bar x_3(t)
		\end{cases}
	\end{align*}
	where $\bar x_3(0) \sim \mathcal{N}(0,\begin{bsmallmatrix} 4 & 0 & 0\\ 0 & 3 & 0 \\ 0 & 0 & 5 \end{bsmallmatrix})$ and $v_3 \sim \mathcal{N}(0,2.05)$.
	
	For $\bar \M_3$, no controller $\bar \C_3$ exist such that $\Pr_{\bar \C_3 \times \bar \M_3}(\mathcal{Z}_\phi) \geq 0.95$, since maximizing the probability satisfaction of the specification $\phi$ leads to the controller $\bar \C_3: \bar u_3(t)=\begin{bmatrix} 0 & -1 & 0 \end{bmatrix} \bar x_3(t)$, resulting in $\bar z_3(t) \sim \mathcal{N}(0,2.05)$. Using the cumulative distribution function again, we can numerically compute that $\Pr_{\bar \C_3 \times \bar \M_3}(\mathcal{Z}_\phi)= 1.543\text{e-}29$. This means no controller for $\bar \M_{3}$ can enforce specification $\phi$. However, previously, we have shown how to enforce this specification for a more complex abstract model. Hence, examples \ref{Exa:1} and \ref{Exa:3} clearly illustrate that while both abstract models simplify the stochastic control synthesis problem, oversimplification may prevent us from finding an adequate controller. \QED
	
\end{example}


\section{Conclusion} \label{Sec:Con}

In this paper, we proposed a model reduction technique on the noise disturbance, whereby a simpler representation of the original dynamics is obtained by means of reducing state information, which we called knowledge filtering, and via optimal Kalman filtering. We introduce the abstraction procedure, which, under some technical conditions, may lead to a time-invariant abstract model. Our controller refinement algorithm is constructive and allows for the design of correct-by-design controllers on the original dynamics via correct-by-design controllers on the abstract model. We finished the paper by illustrating the abstraction procedure and the controller refinement, by showing how our reduction method can achieve model reduction on the state space, which existing methods, such as those employing simulation relations, cannot realize, and by examining the completeness of our technique.


\bibliographystyle{IEEEtranS}
\bibliography{ArticleCDC.bib}


\if\mycmd0
\section*{Appendix}

\noindent \textit{Additional information on the derivations of \eqref{InnPro} and \eqref{CurInnPro}. }

The innovation processes are obtained by simple substitution of the equations \eqref{KalFilEq} and the innovation \eqref{Eq:Inn}. This is trivial in the case of the a priori innovation process. For the a posteriori innovation process, two additional details will be given. The initial state distribution $\bar x(0)$ is obtained from \eqref{KalMeaUpdSta}, $y(0)=C x(0)$, $ x(0) \sim \mathcal{N}(\mu_0,\Sigma_0)$ and $K(0)= \Sigma_0C^T[C\Sigma_0C^T]^{-1}$, resulting in $\bar x(0)=\mu_0+K(0)(y(0)-C\mu_0)$. Calculating mean and variance while also noticing that $x(0)$ is Gaussian, results in the Gaussian distribution of $\bar x(0)$. The observation output $y(t)$ is obtained from substituting \eqref{KalMeaUpdSta} into \eqref{Eq:Inn} resulting in $y(t) = C\bar x(t)- CK(t)v(t) + v(t)$. Since $CK(t)=I$, it follows that $y(t)=C\bar x(t)+(I-CK(t))v(t)=C\bar x(t)$. Alternatively, notice that $\bar x(t)$ contains information about the measured output $y(t)$. Intuitively it makes sense that if the original measurement is used to calculate the expected observation output, the expected observation output will be equivalent to the original measurement.
\bigskip

\begin{lemma}[Theorem \ref{Thm:main}] \label{Thm:PerOut}
	$\M$ and $\bar \M$ have the same performance output distribution $\pmb z$ and $\pmb{\bar{z}}$ when given the same input sequence.
\end{lemma}

\begin{proof}
	The proof follows in 3 steps.
	
	\textbf{Step 1:} $\M$ and $\M_{\text{Obs}}$ have identical performance output distributions when given the same input sequence. This follows directly from $NC=H$. 
	
	\textbf{Step 2:} $\M_{\text{Obs}}$ and $\hat \M$ have the same performance output distributions $\mathbf{z}$ and $\hat{\mathbf{z}}$ when given the same input sequence. To prove this statement, consider the proof of \cite[Prop 8.4.3]{VanS2021}. There are three differences between \cite[Prop 8.4.3]{VanS2021} and the setting in this paper, them being the additional input and output in $\M_{\text{Obs}}$ and $\hat \M$, and the absence of output noise in $\M_{\text{Obs}}$. The former two, however, do not influence the result, as can be observed from comparing \cite[Theorem 8.3.2 \& Theorem 14.4.2]{VanS2021} with the former being the main contributor to the proof of \cite[Prop 8.4.3]{VanS2021}. On the other hand, the exclusion of output noise in $\M_{\text{Obs}}$ is already resolved by the assumption that $C\Sigma_0C^T \succ 0$ and $CQ_wC^T  \succ 0$.
	
	\textbf{Step 3:} $\hat \M$ and $\bar \M$ have the same performance output distributions $\hat{\mathbf{z}}$ and $\bar{\mathbf{z}}$ when given the same input sequence. Since the innovation processes can be obtained from one another by direct substitution of the Kalman filter equations \eqref{KalFilEq} and the innovation \eqref{Eq:Inn}, without changing their performance outputs, both have the same performance outputs $\hat{\mathbf{z}}$ and $\bar{\mathbf{z}}$ when given the same input sequence.
	
	These steps together prove the lemma.
\end{proof}
\bigskip

\noindent \textit{Additional information on the derivation of $\bar \M^*$. }

To obtain abstract model $\bar \M^*$, we need to calculate the a posteriori innovation process while assuming an additional measurement $\tilde y$ is taken at time $t=0$. To accomplish this, we first calculate the a posteriori state estimate $x_K(0|0)=\mathbb{E}(x(0)|y(0),\tilde y)$ and the a posteriori state variance $P(0|0)=\text{Var}(x(0)|y(0),\tilde y)$, to afterwards compute the a posteriori innovation process by utilizing the Kalman filter equations \eqref{KalFilEq}, excluding \eqref{KalFilIni}.

Remember $x(0) \sim \mathcal{N}(\mu_0,\Sigma_0)$, $y(0)=Cx(0)$ and $\tilde y=x(0)+\tilde w$ with $\tilde w \sim \mathcal{N}(0,R)$ and $R=(X^{-1}-\Sigma_0^{-1})^{-1}$. To compute both a posteriori quantities of $x(0)$, we utilize the following proposition derived from \cite[Section 3.1]{Ande2012}.
\begin{proposition}
	If $X$ and $Y$ are jointly Gaussian, with $Z=[X^T \ Y^T]^T$ possessing mean and covariance
	\begin{equation*}
		\begin{bmatrix}
			\mu_x\\
			\mu_y
		\end{bmatrix} \text{ and } 
		\begin{bmatrix}
			\Sigma_{xx} & \Sigma_{xy}\\
			\Sigma_{yx} & \Sigma_{yy}
		\end{bmatrix},
	\end{equation*}
	 then X, conditioned on the information that $Y=y$, is Gaussian, with mean and variance
	 \begin{align*}
	 	\mathbb{E}(X|Y=y)&=\mu_x +\Sigma_{xy}\Sigma_y^{-1}(y-\mu_y),\\ \text{Var}(X|Y=y)&=\Sigma_{xx}-\Sigma_{xy}\Sigma_y^{-1}\Sigma_{yx}.
	 \end{align*}
\end{proposition}
Utilizing the above proposition, we first calculate the following quantities 
\begin{align*}
	\mathbb{E}(x(0)|\tilde y)&=\mu_0+\Sigma_0(\Sigma_0+R)^{-1}(\tilde y-\mu_0),\\
	\text{Var}(x(0)|\tilde y)&=\Sigma_0-\Sigma_0(\Sigma_0+R)^{-1}\Sigma_0,\\
	\mathbb{E}(y(0)|\tilde y)&= C\mathbb{E}(x(0)|\tilde y),\\
	\text{Var}(y(0)|\tilde y)&=C\text{Var}(x(0)|\tilde y)C^T.
\end{align*}
Utilizing the Woodbury matrix identity, we identify that
\begin{align*}
	R &= (X^{-1}-\Sigma_0^{-1})^{-1} \\
	R& = (\Sigma_0^{-1}-\Sigma_0^{-1}X\Sigma_0^{-1})^{-1}-\Sigma_0 \\
	(R+\Sigma_0)^{-1}  &= \Sigma_0^{-1}-\Sigma_0^{-1}X\Sigma_0^{-1}  \\
	\Sigma_0(R+\Sigma_0)^{-1}\Sigma_0 & = \Sigma_0-X \\
	X &= \Sigma_0 - \Sigma_0(R+\Sigma_0)^{-1}\Sigma_0,
\end{align*}
are all equivalent. Let $\mathbb{E}(x(0)|\tilde y)=\bar \mu_0$ and notice that $\text{Var}(x(0)|\tilde y)=X$ and $\text{Cov}(x(0),y(0)|\tilde y)=XC^T$. We can now directly compute $x_K(0|0)$ and $P(0|0)$, again by utilizing the above proposition. The result is given by
\begin{subequations}
	\begin{align}
			x_K(0|0)&=\bar \mu_0+XC^T(CXC^T)^{-1}(y(0)-C\bar \mu_0),\\
			P(0|0)&=X-XC^T(CXC^T)^{-1}CX.
		\end{align}
\end{subequations}
Notice that $x_K(0|0) \sim \mathcal{N}(\mu_0, \Sigma_0-P(0|0))$ and $P(1|0)=X$, the latter of which implies that the a posteriori innovation process will be time-invariant. The a posteriori innovation process can now be obtained from the Kalman filter equations \eqref{KalFilEq}, excluding \eqref{KalFilIni}.
\else

\fi
\end{document}

%% file: commands.tex
\newcommand{\Borel}{\mathcal{B}}
\renewcommand{\Pr}{\mathbb{P}}
\newcommand{\B}{\mathcal{B}}

\newcommand{\C}{\mathbf{C}}
\newcommand{\M}{\mathbf{M}}
\newcommand{\X}{\mathbb{X}}

